\documentclass[a4paper,12pt]{article}
\usepackage{jheppub,esint,shuffle,psfrag}
\usepackage[utf8]{inputenc}

\usepackage{epsfig,amssymb,amsmath,psfrag,subfigure,rotate,color,wasysym,xcolor}
\usepackage[bbgreekl]{mathbbol}

\allowdisplaybreaks 
\newcommand{\insertfig}[2]{\includegraphics[width=#1cm]{#2}}

\DeclareSymbolFontAlphabet{\mathbbm}{bbold}
\DeclareSymbolFontAlphabet{\mathbb}{AMSb}%

\def\XXint#1#2#3{{\setbox0=\hbox{$#1{#2#3}{\int}$ }
\vcenter{\hbox{$#2#3$ }}\kern-.6\wd0}}

\def \be  {\begin{equation}}
\def \ee  {\end{equation}}
\def \ba  {\begin{eqnarray}}
\def \ea  {\end{eqnarray}}
\def \baa {\begin{eqnarray*}}
\def \eaa {\end{eqnarray*}}

\def \lab #1 {\label{#1}}

\newcommand\re[1]{(\ref{#1})}
\def\d{\hbox{{d}\kern-.20em\hbox{l}}}

\def \matrix #1 {\left(\begin{array}{cc} #1 \end{array}\right)}

\def \tr {\mathop{\rm tr}\nolimits}

\def \e  {\mathop{\rm e}\nolimits}

\newcommand{\bit}[1]{\mbox{\boldmath$#1$}}
\def\1{\hbox{{1}\kern-.25em\hbox{l}}}

\newcommand{\ft}[2]{{\textstyle\frac{#1}{#2}}}
\makeatletter

\input pdf-trans
\newbox\qbox
\def\usecolor#1{\csname\string\color@#1\endcsname\space}
\newcommand\bordercolor[1]{\colsplit{1}{#1}}
\newcommand\fillcolor[1]{\colsplit{0}{#1}}
\newcommand\outline[1]{\leavevmode%
  \def\maltext{#1}%
  \setbox\qbox=\hbox{\maltext}%
  \boxgs{Q q 2 Tr \thickness\space w \fillcol\space \bordercol\space}{}%
  \copy\qbox%
}
\makeatother
\newcommand\colsplit[2]{\colorlet{tmpcolor}{#2}\edef\tmp{\usecolor{tmpcolor}}%
  \def\tmpB{}\expandafter\colsplithelp\tmp\relax%
  \ifnum0=#1\relax\edef\fillcol{\tmpB}\else\edef\bordercol{\tmpC}\fi}
\def\colsplithelp#1#2 #3\relax{%
  \edef\tmpB{\tmpB#1#2 }%
  \ifnum `#1>`9\relax\def\tmpC{#3}\else\colsplithelp#3\relax\fi
}
\bordercolor{black}
\fillcolor{white}
\def\thickness{.3}

\def\1{\mathbbm{1}}

\title{Three-leg form factor on Coulomb branch}
 
\author[a]{A.V.~Belitsky,}
\author[b]{L.V.~Bork,}
\author[a]{J.M.~Grumski-Flores,}
\author[c]{V.A. Smirnov}
\affiliation[a] {Department of Physics, Arizona State University,  Tempe, AZ 85287-1504, USA}  
\affiliation[b]{Institute for Theoretical and Experimental Physics, 117218 Moscow, Russia}
\affiliation[]{The Center for Fundamental and Applied Research, 127030 Moscow, Russia}
\affiliation[c]{Skobeltsyn Institute of Nuclear Physics, Moscow State University 119992 Moscow, Russia}
\affiliation{Moscow Center for Fundamental and Applied Mathematics 119992 Moscow, Russia} 
  
 \abstract
{We study the form factor of the lowest component of the stress-tensor multiplet away from the origin of the moduli space
in the spontaneously broken, aka Coulomb, phase of the maximally supersymmetric Yang-Mills theory for decay into three
massive W-bosons. The calculations are done at two-loop order by deriving and solving canonical differential equations
in the asymptotical limit of nearly vanishing W-masses. We confirm our previous findings that infrared physics of `off-shell
observables' is governed by the octagon anomalous dimension rather than the cusp. In addition, the form factor in question 
possesses a nontrivial remainder function, which was found to be identical to the massless case, upon a proper subtraction 
of infrared logarithms (and finite terms). However, the iterative structure of the object is more intricate and is not simply 
related to the previous orders in coupling as opposed to amplitudes/form factors at the origin of the moduli space.}

\begin{document}

\maketitle
\flushbottom
\setcounter{footnote} 0

\section{Introduction}
\label{s1}

The so-called Coulomb branch of the spontaneously broken maximally supersymmetric Yang-Mills (sYM) theory \cite{Alday:2009zm} is a 
natural laboratory to study off-shell amplitudes and form factors in four-dimensional gauge theories. Endowing some (or all) scalars of the 
model with vacuum expectation values, one can adjust their values in a way as to yield matrix elements which possess massive external 
states, i.e., W-bosons, Higgs-like scalars etc., but only massless excitations propagating in quantum loops.

The $\mathcal{N} = 4$ model away from the origin of the moduli space can naturally be obtained from a generalized form of the dimensional 
reduction \cite{Selivanov:1999ie,Boels:2010mj,Craig:2011ws} akin to the original one used to discover its Lagrangian in the first place from 
the ten-dimensional $\mathcal{N} = 1$ sYM \cite{Brink:1976bc,Gliozzi:1976qd}, or the six-dimensional $\mathcal{N} = (1,1)$ sYM (see, e.g., 
\cite{Bern:2010qa}). Instead of setting the extra-dimensional, i.e., $D>4$, components of momenta to zero, one can trade them in lieu of 
the scalars' moduli, i.e., vacuum averages of $D>4$ components of the ten-dimensional gauge field. The 16 supercharges remain unbroken 
in this phase, but the supersymmetric algebra gets a central extension 
with BPS charges induced by nonvanishing masses, and thus the theory shares a gamut of properties of its conformal sibling. One-loop analyses 
demonstrated that the Coulomb branch scattering amplitudes obey a no-triangle rule\footnote{Bubbles and tadpoles are excluded form the 
get-go based on their poor ultraviolet properties.}, thus enjoying only boxes in their integral expansion \cite{Schabinger:2008ah,Boels:2010mj} 
at a generic point of the moduli space, --- a generalization of the background-field gauge proof from Refs.\ \cite{Gates:1983nr,Bern:1994zx} 
applicable to off-shell massless amplitudes. Further, there are no rational terms as well \cite{Badger:2008cm,Boels:2010mj} and, therefore, 
integrands on the Coulomb branch are cut-constructible \cite{Bern:1994cg}. Making use of this latter property, a proof of the dual conformal 
invariance of massive loop integrals from a six-dimensional viewpoint was elucidated in Refs.\ \cite{Bern:2010qa,Dennen:2010dh}. Also  
correctness of the four-leg amplitudes at four-loop order (including nonplanar contributions) \cite{Bern:2010qa} was demonstrated to expressions 
built using solely four-dimensional momenta in the cuts \cite{Bern:2010tq} by lifting four-dimensional inner products of momenta up to six 
dimensions.

The above higher-dimensional perspective provides a natural bridge between the dimensionally regularized theory and its massive version 
to tame infrared divergences in scattering amplitudes and form factors in a gauge invariant manner. Their explicit structure for the four-gluon 
amplitude and the Sudakov form factor was deduced at up to three-loop level \cite{Alday:2009zm,Henn:2010bk,Henn:2010ir,Henn:2011by} by 
promoting massless integral bases constructed in four dimensions to involve massive propagators only around graphs' periphery. Infrared 
structure was shown to be in accord with the well-known conformal phase of  $\mathcal{N} = 4$ sYM in $D = 4 - 2 \varepsilon$ \cite{Bern:2005iz} 
(see Refs.\ \cite{Mueller:1979ih,Magnea:1990zb,Sterman:2002qn}, for earlier QCD studies) for a minor difference in kinematically-independent 
contributions and in compliance with a common wisdom that infrared properties of gauge theories are encoded in the so-called cusp anomalous 
dimension \cite{Polyakov:1980ca,Korchemsky:1987wg}.

The situation drastically changes, however, when all internal propagators are left massless, but the external legs are kept massive, or 
off-shell, as we will refer to them hereafter. Four- \cite{Caron-Huot:2021usw} and five-leg \cite{Bork:2022vat} W-boson amplitudes as well 
as the two--W-boson Sudakov form factor \cite{Belitsky:2022itf,Belitsky:2023ssv} enjoyed the same recurrent feature in variance to the 
naive expectation: the infrared logarithms are governed by an exponent different from the cusp anomalous dimension. Instead they exhibit 
dependence on the so-called octagon anomalous dimension which made its debut in completely different circumstances: the light-cone 
limit of correlation functions of infinitely heavy BPS operators \cite{Coronado:2018cxj,Belitsky:2019fan} and the near-origin asymptotics of 
the six-gluon remainder function \cite{Basso:2020xts}.

In the current paper, we continue our exploration of the Coulomb branch by addressing a more involved quantity, the three--W-boson form 
factor $\mathcal{F}_3$ of the lowest component of the stress tensor multiplet
\begin{align}
\label{MatrixElement}
\int d^4 x \, {\rm e}^{- i q \cdot x} \langle p_1, p_2, p_3 | \tr \phi_{12}^2 (x) | 0 \rangle
=
(2 \pi)^4 \delta^{(4)} (q - p_1 - p_2 - p_3) \mathcal{F}_3 
\, .
\end{align}
Here we explicitly extracted the energy-momentum conserving delta function.
This is the simplest `observable' which possesses nontrivial remainder function after factoring out infrared divergences \cite{Brandhuber:2012vm}. 
In the conformal phase of the theory, it was bootstrapped to a staggering eight-loop order \cite{Dixon:2020bbt,Dixon:2022rse} using 
techniques adopted from scattering amplitudes \cite{Dixon:2011pw}. Our goal will be much more modest: we will calculate its off-shell version 
at two loops. The incentive for our analysis is multifold. First, we would like to confirm the octagon anomalous dimension as the Sudakov
exponent of `off-shell observables'. Second, we will establish similarities/differences to the iterative structure of the form factor with increased
perturbative order compared to its conformal analogue. Third, given that the infrared logarithms are different in the on- and off-shell cases, will the
remainder functions differ as well?

Our subsequent presentation will be organized as follows. In the next section, we set up our notations. Then, in Sect.\ \ref{1LoopSection}, 
we perform the one-loop calculation, which is then followed by two loops in Sect.\ \ref{2LoopSection}. The only graph that was not touched
upon in the existing literature corresponds to the tri-pentagon. So we perform its calculation from scratch in Sect.\ \ref{TriPentagonSection}. 
It is then followed by all other contributing graphs. In Sect.\ \ref{SumSection}, we add them up and use symbol analysis to simplify the sum and 
uncover the structure of the form factor at two-loop order. Finally, we conclude.

\section{Setting up conventions}

The form factor of three W-bosons contains an overall prefactor encoding polarization dependence of the external states. We
will not be interested in it in what follows and thus introduce the ratio function
\begin{align}
F_3 \equiv \mathcal{F}_3/\mathcal{F}_{3, {\rm tree}}
\, .
\end{align} 
$F_3$ depends on three invariants $s_{ij}$ and the off-shellnesses  of the W-legs, which will be taken to have the same value $\mu$,
\begin{align}
s_{ij} \equiv (p_i + p_j)^2 \, , \quad p_i^2 \equiv - \mu
\, .
\end{align}
These are linearly dependent, however,
\begin{align}
\label{LDcondition}
s_{12} + s_{23} + s_{31} = q^2 - 3 \mu
\, .
\end{align}
Since the form factor is a homogeneous function of these kinematical variables, one can set one scale to one, e.g., $q^2 = -1$ below. 
Equivalently this can be done by introducing Mandelstam-like variables and the `mass parameter' $m$
\begin{align}
u = s_{12}/q^2
\, , \quad
v = s_{23}/q^2
\, , \quad
w = s_{31}/q^2
\, , \quad
m = - \mu/q^2
\, ,
\end{align}
and ignore the overall mass scale $q^2$: $F_3$ is dimensionless.

$F_3$ admits perturbative series expansion in the gauge coupling $g^2_{\scriptscriptstyle\rm YM}$ accompanied at each order by the 
number of colors $N$ (in the planar limit), allowing us to introduce
\begin{align}
g^2 \equiv \frac{g^2_{\scriptscriptstyle\rm YM} N}{(4 \pi)^2} (4 \pi {\rm e}^{- \gamma_{\rm E}})^{\varepsilon}
\, ,
\end{align}
which comes hand-in-hand with a measure of the dimensionally-regularized loop momentum integrals,
\begin{align}
(\mu_{\scriptscriptstyle\rm DR}^2 {\rm e}^{\gamma_{\rm E}})^{\varepsilon}  \int \frac{d^{D} \ell}{i \pi^{D/2}}
\, ,
\end{align}
in $D = 4 - 2 \varepsilon$. We will dwell on the necessity to deal with the space-time away from $D=4$, even though we already have an 
infrared regulator $m$, when it becomes indispensable at two-loop order. 

Thus, we have to the lowest two orders
\begin{align}
F_3 = 1 + g^2 F^{(1)}_3 + g^4 F^{(2)}_3 + \dots
\, ,
\end{align}
where $F^{(i)}_3$ are given by linear combinations of one- and two-loop integrals for $i=1,2$, respectively. Instead of using Feynman 
rules of the Coulomb phase of $\mathcal{N} = 4$ sYM in order to find the latter, we will employ, as was advocated in the introduction, the 
connection between the spontaneously broken phase in $D=4$ and higher-dimensional theory with exact gauge symmetry to recycle 
generalized unitarity analyses from Refs.\ \cite{Brandhuber:2010ad,Bork:2011cj} and \cite{Brandhuber:2012vm}, to ascertain integral 
families defining the one and two-loop integrands, respectively. All calculations will be done in the limit $m \to 0$, i.e., they will be valid 
up to power corrections in $m$.

\section{One loop}
\label{1LoopSection}

%%%%%%%%%%%%%%%%%%%%%%%%%%%%%%%%%%%%%%%%%%%%%%%%%%%%%%%%%%%%%%%%%%%%%
%            Figure
%%%%%%%%%%%%%%%%%%%%%%%%%%%%%%%%%%%%%%%%%%%%%%%%%%%%%%%%%%%%%%%%%%%%%
\begin{figure}[t]
\begin{center}
\mbox{
\begin{picture}(0,120)(180,0)
\put(0,0){\insertfig{12}{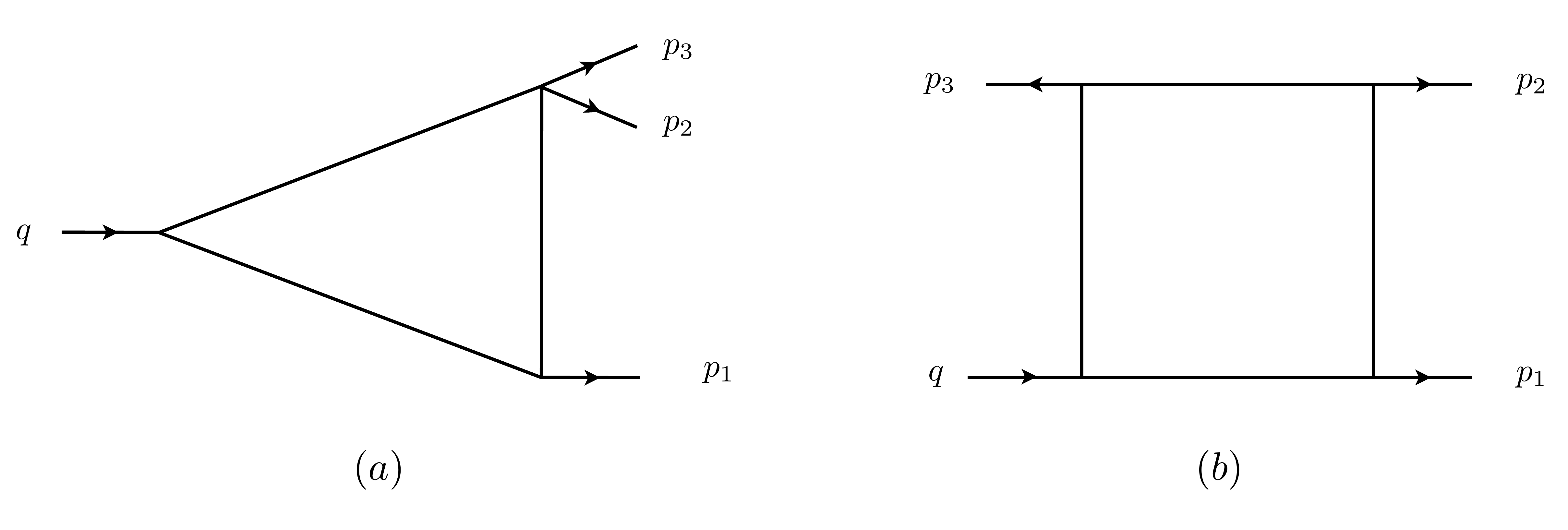}}
\end{picture}
}
\end{center}
\caption{\label{fig1loop} Graphs contributing to the one-loop form factor $F_3^{(1)}$.}
\end{figure}
%%%%%%%%%%%%%%%%%%%%%%%%%%%%%%%%%%%%%%%%%%%%%%%%%%%%%%%%%%%%%%%%%%%%%

Without further ado, let us start our analysis with the one-loop form factor. At this order, $F_3^{(1)}$ receives the following expansion in 
terms of the triangle ${\rm Tri}$ and box {\rm Box} integrals \cite{Brandhuber:2010ad,Bork:2011cj}
\begin{align}
\label{1loopFF}
F_3^{(1)}
&= 
\sum_{n = 0}^2 \mathbb{P}^n 
\left[
(s_{12} + s_{13}) {\rm Tri} (p_1, p_2 + p_3) + \frac12 s_{12} s_{23} {\rm Box} (p_1, p_2, p_3)
\right]
\, ,
\end{align}
shown in Fig.\ \ref{fig1loop}. In this equation, we introduced an operator $\mathbb{P}$ that shifts momentum indices of any function to its 
right by one
\begin{align}
\mathbb{P} f_{ij \dots} = f_{i+1 ({\rm mod \, 3}), j+1 ({\rm mod \, 3}) \dots}
\, ,
\end{align}
modulo 3, which imposes periodicity. The Mandelstam-like variables then transform as $\mathbb{P} (u,v,w) = (v,w,u)$. Both 
integrals in Eq.\ \re{1loopFF} can immediately be expressed in terms of the Davydychev-Ussyukina function $\Phi_1 (x,y)$ 
\cite{Usyukina:1992jd,Usyukina:1993ch},
\begin{align}
\label{BoxFunct}
\Phi_\ell(x,y)=-\sum_{j=\ell}^{2\ell}\frac{j!(-1)^{j}\log^{2 \ell - j}\left( \frac{y}{x} \right)}{\ell! (j - \ell)! (2 \ell - j)!}
\frac{\mbox{Li}_{j} \big(-(\rho x)^{-1}\big) - \mbox{Li}_{j}\big( -(\rho y)^{+1}\big)}
{\lambda} 
\, ,
\end{align}
where $\rho$ and $\lambda$ are functions of $x$ and $y$,
\begin{align}
\lambda(x,y)=[(1-x-y)^2-4xy]^{1/2} \, , \qquad \rho(x,y)=2 [1-x-y-\lambda(x,y)]^{-1} \, ,
\end{align}
via
\begin{align}
{\rm Tri} (p_1, p_2 + p_3) 
&= \Phi_1(m,v)/(1 - v)
\\
{\rm Box} (p_1, p_2, p_3)
&= \Phi_1 \left(m^2/(u v), m/(u v)\right)/(u v)
\, .
\end{align}
Their small-mass expansion yields the following expressions for the triangle and the box 
\begin{align}
{\rm Tri} (p_1, p_2 + p_3) 
&
=
- \frac{\log m \log v + 2 {\rm Li}_2 (1-v)}{1-v}
\, , \quad\\
{\rm Box} (p_1, p_2, p_3)
&
=
- \frac{2 \log^2 m - 2 \log m \log (u v) + \log^2 (u v) + 2 \zeta_2}{u v}
\, ,
\end{align}
where we used the condition \re{LDcondition}, up to terms vanishing as a power of $m$. Adding all contributions up, we find
\begin{align}
F_3^{(1)} 
&
= - \log^2 \frac{m}{u} - \log^2 \frac{m}{v} - \log^2 \frac{m}{w}
\nonumber\\
&
- \log u \log v  - \log v \log w  - \log w \log u 
\nonumber\\
&
- 2  {\rm Li}_2 (1-u) - 2  {\rm Li}_2 (1-v) - 2  {\rm Li}_2 (1-w)
- 3 \zeta_2 
\, .
\end{align}
It is instructive to compare this results to the conformal case, calculated within dimensional regularization (or rather reduction),
\cite{Brandhuber:2010ad},
\begin{align}
\label{1loopDR}
F_3^{(1)} (\varepsilon) 
&
= 
- \frac{1}{\varepsilon^2}  
\left[
\left( \frac{- \mu_{\scriptscriptstyle\rm DR}^2}{u}\right)^\varepsilon 
+ 
\left( \frac{- \mu_{\scriptscriptstyle\rm DR}^2}{v}\right)^\varepsilon 
+ 
\left( \frac{- \mu_{\scriptscriptstyle\rm DR}^2}{w}\right)^\varepsilon 
\right]
\nonumber\\
&
- \log u \log v  - \log v \log w  - \log w \log u 
\nonumber\\
&
- 2  {\rm Li}_2 (1-u) - 2  {\rm Li}_2 (1-v) - 2  {\rm Li}_2 (1-w)
+ 
\frac{9}{2} \zeta_2 
\, .
\end{align}
We observe that the finite parts are identical in the two cases, except for the coefficient of $\zeta_2$. When Eq.\ \re{1loopDR} expanded in 
the Laurent series, the coefficient of the double logarithms of $\mu_{\scriptscriptstyle\rm DR}^2/(u,v,w)$ are half of the off-shell case, as anticipated. 
This is the well-known doubling phenomenon observed back in the early days of QED \cite{Sudakov:1954sw,Jackiw:1968zz} and 
well-understood by now as a result of an extra, the so-called ultra-soft, region \cite{Fishbane:1971jz,Mueller:1981sg,Korchemsky:1988hd} 
of loop momentum producing leading effects on par with other regimes present in both.

\section{Two loops}
\label{2LoopSection}

%%%%%%%%%%%%%%%%%%%%%%%%%%%%%%%%%%%%%%%%%%%%%%%%%%%%%%%%%%%%%%%%%%%%%
%            Figure
%%%%%%%%%%%%%%%%%%%%%%%%%%%%%%%%%%%%%%%%%%%%%%%%%%%%%%%%%%%%%%%%%%%%%
\begin{figure}[t]
\begin{center}
\mbox{
\begin{picture}(0,370)(220,0)
\put(0,0){\insertfig{15}{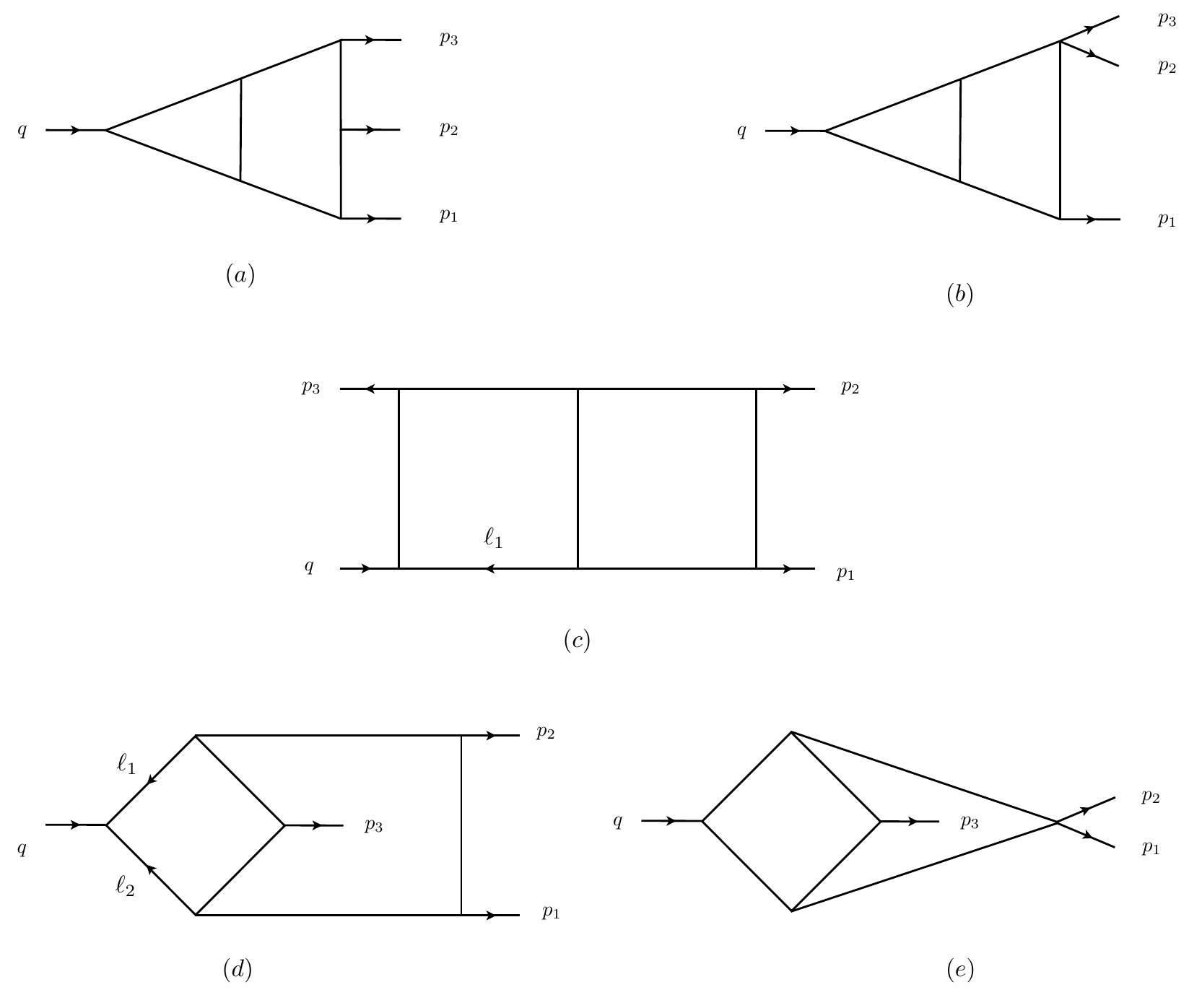}}
\end{picture}
}
\end{center}
\caption{\label{fig2loop} Graphs contributing to the two-loop form factor $F_3^{(2)}$. The integrands, built from product of 
propagators read off from these diagrams, are accompanied by numerators according to Eqs.\ \re{TriPent}--\re{NTriBox}.}
\end{figure}
%%%%%%%%%%%%%%%%%%%%%%%%%%%%%%%%%%%%%%%%%%%%%%%%%%%%%%%%%%%%%%%%%%%%%
 
\begin{figure}[t]
\begin{center}
\mbox{
\begin{picture}(0,200)(80,0)
\put(0,0){\insertfig{5}{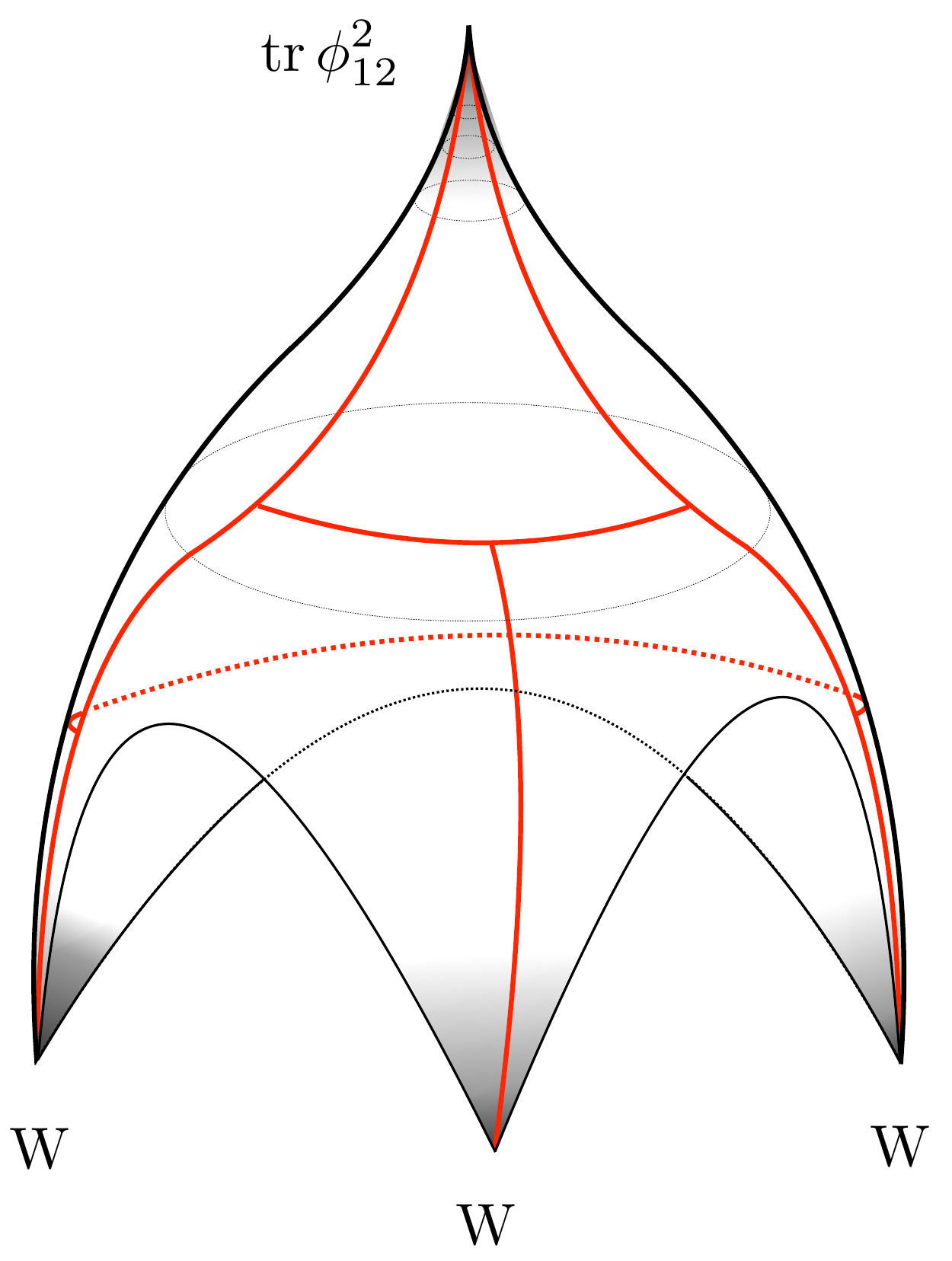}}
\end{picture}
}
\end{center}
\caption{\label{figWS} World-sheet perspective of the three-leg form factor and the non-planar graph from Fig.\ \ref{fig2loop} (d) overlayed 
on it: it demonstrates why it produces contribution of leading order in color.}
\end{figure}

We now proceed to the two loop calculation. The integrands for the form factor $F_3^{(2)}$ were constructed in Ref.\ \cite{Brandhuber:2012vm}
using (generalized) unitarity cut technique. With a slight change of the nomenclature compared to \cite{Brandhuber:2012vm}, the  relevant graphs 
shown in Fig.\ \ref{fig2loop} generate the following integrals\footnote{Here and below, we set the mass scale of dimensional regularization to one, 
$\mu_{\scriptscriptstyle\rm DR}^2 = 1$}
\begin{align}
\label{TriPent}
{\rm TriPent} (p_1, p_2, p_3)
&
=
{\rm e}^{2 \varepsilon \gamma_{\rm E}} \int \frac{d^{D} \ell_1}{i \pi^{D/2}} \int \frac{d^{D} \ell_2}{i \pi^{D/2}}
\frac{q^2 s_{12} s_{23}}{{\rm denom}_{(a)}}
\, , \\
\label{TriBox}
{\rm TriBox} (p_1, p_2 + p_3)
&
=
{\rm e}^{2 \varepsilon \gamma_{\rm E}}  \int \frac{d^{D} \ell_1}{i \pi^{D/2}}  \int \frac{d^{D} \ell_2}{i \pi^{D/2}}
\frac{q^2  [s_{12} + s_{31}] }{{\rm denom}_{(b)}}
\, , \\
\label{Dbox}
{\rm DBox} (p_1, p_2, p_3)
&
=
{\rm e}^{2 \varepsilon \gamma_{\rm E}} \int \frac{d^{D} \ell_1}{i \pi^{D/2}} \int \frac{d^{D} \ell_2}{i \pi^{D/2}}
\frac{s_{12} [ s_{31} \, \ell_1 \cdot p_1 - s_{23} \, \ell_1 \cdot p_2]}{{\rm denom}_{(c)}}
\, , \\
\label{Nbox}
{\rm NBox} (p_1, p_2, p_3)
&
=
{\rm e}^{2 \varepsilon \gamma_{\rm E}}  \int \frac{d^{D} \ell_1}{i \pi^{D/2}} \int \frac{d^{D} \ell_2}{i \pi^{D/2}}
\frac{s_{12} [ \ft12 s_{23} s_{31} -  s_{23} \ell_1 \cdot p_2 - s_{31} \, \ell_2 \cdot p_1]}{{\rm denom}_{(d)}}
\, , \\
\label{NTriBox}
{\rm NTriBox} (p_1 + p_2, p_3)
&
=
{\rm e}^{2 \varepsilon \gamma_{\rm E}}  \int \frac{d^{D} \ell_1}{i \pi^{D/2}} \int \frac{d^{D} \ell_2}{i \pi^{D/2}}
\frac{\ft12 q^2 [s_{23} + s_{31}]}{{\rm denom}_{(e)}}
\, , 
\end{align}
where the denominator structure can readily be read off from the corresponding graphs. In terms of these integrals, the two-loop 
form factor is given by the expression
\begin{align}
\label{2loopFF}
F_3^{(2)}
=
\sum_{n = 0}^2 \mathbb{P}^n
\bigg[
&
{\rm TriBox}  (p_1, p_2 + p_3) 
+ {\rm TriBox}  (p_3, p_1 + p_2) + {\rm TriPent} (p_1, p_2, p_3)
\nonumber\\
&
+
{\rm DBox} (p_1, p_2, p_3) + {\rm DBox} (p_3, p_2, p_1) + {\rm NBox} (p_1, p_2, p_3)
\nonumber\\
&
\qquad\qquad\qquad\qquad\,
+
{\rm NTriBox} (p_1 + p_2, p_3)
\bigg]
\, .
\end{align}
Notice that starting from this order, there are non-planar graphs which are leading order in color, i.e., Fig.\ \ref{fig2loop} $(d)$ and $(e)$. 
The reason for this is that the operator $\tr \phi_{12}^2$ is a singlet with respect to the SU$(N)$ and thus does not `participate' in 
color traces. It becomes quite obvious from the world-sheet perspective of the matrix element \re{MatrixElement} demonstrated
in Fig.\ \ref{figWS} where the operator corresponds to the closed string state, while the W-bosons to the open ones.

%%%%%%%%%%%%%%%%%%%%%%%%%%%%%%%%%%%%%%%%%%%%%%%%%%%%%%%%%%%%%%%%%%%%%
%            Figure
%%%%%%%%%%%%%%%%%%%%%%%%%%%%%%%%%%%%%%%%%%%%%%%%%%%%%%%%%%%%%%%%%%%%%
%%%%%%%%%%%%%%%%%%%%%%%%%%%%%%%%%%%%%%%%%%%%%%%%%%%%%%%%%%%%%%%%%%%%%

Out of all the contributions in Eq.\ \re{2loopFF}, a truly new integral, which was not addressed in existing literature, is the tri-pentagon,
Fig.\ \ref{fig2loop} $(a)$. So we start with its analysis first in the next section.

\subsection{Tri-pentagon}
\label{TriPentagonSection}

Let us begin with the construction of the canonical basis for the tri-pentagon family, see Fig.\ \ref{fig2loop} $(a)$, by routing the loop 
momenta $\ell_1$ and $\ell_2$ according to the following definitions of propagator denominators $D_{i}$ ($i = 1, \dots, 7$) and 
irreducible scalar products $D_8$ and $D_9$,
\begin{align}
&D_1 = - \ell_1^2
\, ,
& 
&D_2 = - (\ell_1 + p_1)^2
\, ,
&
&D_3 = - (\ell_1 + p_1 + p_2)^2
\, ,
& 
&D_4 = - (\ell_1 + p_1 + p_2 + p_3)^2
\, , \nonumber\\
&D_5 = - \ell_2^2
\, , 
&
&D_6 = - (\ell_2 - \ell_1)^2
\, ,
& 
&D_7 = - (\ell_2 + p_1 + p_2 + p_3)^2
\, ,
& 
&D_8 = - (\ell_2 + p_1)^2
\, , \nonumber\\
&&
&&
&
D_{9}=-(\ell_2 + p_1 + p_2)^2
\, ,
&
&&
\end{align}
such that
\begin{align}
G_{a_1 a_2 a_3 a_4 a_5 a_6 a_7 a_8 a_9}
\equiv
{\rm e}^{2 \varepsilon \gamma_{\rm E}}  \int \frac{d^{D} \ell_1}{i \pi^{D/2}}  \int \frac{d^{D} \ell_2}{i \pi^{D/2}}
\prod_{i=1}^9 D_i^{- a_1}
\, .
\end{align}
The use of dimensionally regularized integrals is required for proper use of the integration-by-part technique in order to work with 
vanishing surface loop integrals, which are at the heart of the formalism \cite{Chetyrkin:1981qh}. An IBP reduction with the {\tt FIRE} 
code \cite{Smirnov:2008iw,Smirnov:2019qkx,Smirnov:2023yhb} immediately reveals 49 initial Master Integrals (MIs), which are
further reduced to 46 by finding equivalences among them with the {\tt LiteRed} software \cite{Lee:2013mka,Lee:2013mka},
generating thus the primary basis
\begin{align}
\bit{I}
=
\{
&
G_{000111000}, G_{001011000}, G_{010001100}, G_{010011000}, G_{001011100}, G_{001011200}, 
\nonumber\\
&
G_{001111000}, G_{001112000}, G_{010011100}, G_{010011200}, G_{010110100}, G_{010111000},
\nonumber\\
&
G_{010112000}, G_{011001100}, G_{011001200}, G_{011011000}, G_{011012000}, G_{100110100},
\nonumber\\
&
G_{101001100}, G_{101001200}, G_{101010100}, G_{110001100}, G_{110001200}, G_{110010100},
&
\nonumber\\
&
G_{001111100}, G_{010111100}, G_{011011100}, G_{011011200}, G_{011012100}, G_{011012200},
&
\nonumber\\
&
G_{011021100}, G_{012011100}, G_{021011100}, G_{011110100}, G_{011111000}, G_{011112000},
\nonumber\\
&
G_{101011100}, G_{101110100}, G_{110011100}, G_{110110100}, G_{111001100}, G_{111001200},
\nonumber\\
&
G_{111010100}, G_{011111100}, G_{111011100}, G_{111110100}
\}
\, .
\end{align}

Next we turn to the derivation of differential equations for $\bit{I}$ making use of a {\tt FIRE} interface to {\tt LiteRed},
\begin{align}
\label{IniDEs}
\partial_i \bit{I} = \bit{M}_i \cdot \bit{I}
\, ,
\end{align}
in the kinematical invariants $i = u,v,w,m$. The goal is now to convert them to the canonical form \cite{Henn:2013pwa}
\begin{align}
\label{CanDEs}
\partial_i \bit{J} = \varepsilon \bit{A}_i \cdot \bit{J}
\, , \qquad
\varepsilon \bit{A}_i = \bit{T}^{-1} \cdot \bit{M} \cdot \bit{T} - \bit{T}^{-1} \cdot \partial_i \bit{T}
\, , 
\end{align}
with some transformation matrix $\bit{T}$. In fact, what we need is an {\sl asymptotically} canonical basis, which captures
all logarithmically enhanced and constant terms in $m$ as $m$ goes to zero. This can  easily be accomplished by keeping 
track of only singular power-like terms in the `virtuality' matrix 
\begin{align}
\bit{A}_m = \frac{\varepsilon}{m} \bit{A}^0_{m} + O(m^0)
\, ,
\end{align}
as was explained at length in Ref.\ \cite{Belitsky:2023gba}.

Splitting the basis elements of $\bit{I}$ into sectors, we form their linear combinations accompanied by unknown functions of
the Mandelstam-like variables $(u,v,w)$ and fix the former by enforcing the $\varepsilon$-form of the differential equations \re{CanDEs}.
Having fixed the diagonal blocks in this manner, the off-diagonal ones can be constrained by using two available software
packages {\tt Canonica} \cite{Meyer:2016slj,Meyer:2017joq} and {\tt Libra} \cite{Lee:2014ioa,Lee:2020zfb}. To achieve this,
one first transforms the equations to the Fuchsian, i.e., dLog, form followed by factorization of the $\varepsilon$-dependence
into an overall factor \cite{Lee:2014ioa}. {\tt Canonica} is solely based on built-in Mathematica commands and fails to successfully
solve corresponding systems of linear equations. Therefore, we used two strategies in our analysis. One was based exclusively on {\tt Libra}.
However, having constructed canonical form of differential equations, we discovered that five of its elements did not possess uniform
transcendentality\footnote{We would like to thank Johannes Henn for instructive communications on this point.} (UT), namely, $J_i$'s 
with indices $i = 34,43,44,45,46$. So in our attempt to alleviate this problem, we deduced yet another form of the canonical differential 
equations by the combined use of {\tt Canonica} (to bring equations to the Fuchsian form) and {\tt Libra} (for the derivation of the 
$\varepsilon$-form). Though, the basis found was slightly different from the first one, nevertheless the very same five elements 
suffered from the very same problem. Obviously, this was not in any way an obstruction in our subsequent steps of solving theses 
`canonical' equations rather it was merely a nuisance: instead of fixing a set of integration constant of uniform transcendentality at each 
$\varepsilon^n$-order, we had to use a sum of constants of increasing transcendental weight $w_i \leq n$. The asymptotically 
canonical basis, which we used in the explicit iterative solution of the differential equations, is
\begin{align}
\label{TriPentCB}
J_{1} &=\varepsilon^2 (3 m-u-v-w) G_{0 0 0 1 2 2 0 0 0}
\, , \\
J_{2} &=\varepsilon^2 u G_{0 0 1 0 2 2 0 0 0}\, , \\
J_{3} &=\varepsilon^2 v G_{0 1 0 0 0 2 2 0 0}\, , \\
J_{4} &=\varepsilon^2 m G_{0 1 0 0 2 2 0 0 0}\, , \\
J_{5} &=\varepsilon^3 (v+w) G_{0 0 2 0 1 1 1 0 0}\ , , \\
J_{6} &=\varepsilon^2 (u+v+w) ((2\varepsilon-1) G_{0 0 1 0 1 1 2 0 0}+\varepsilon G_{0 0 2 0 1 1 1 0 0})\, , \\
J_{7} &=\varepsilon^2 (2\varepsilon-1) m G_{0 0 2 1 1 1 0 0 0}\, , \\
J_{8} &=\varepsilon^3 (v+w) G_{0 0 1 1 1 2 0 0 0}\, , \\
J_{9} &=\varepsilon^3 (u+w) G_{0 2 0 0 1 1 1 0 0}\, , \\
J_{10} &= \varepsilon^2(1-2\varepsilon) (u+v+w) G_{0 1 0 0 1 1 2 0 0}\, , \\
J_{11} &= \frac{7}{25} \varepsilon^2 (1-2\varepsilon)^2 G_{0 1 0 1 1 0 1 0 0}\, , \\
J_{12} &=\varepsilon^2 (2\varepsilon-1) (3\varepsilon-1) G_{0 1 0 1 1 1 0 0 0}\, , \\
J_{13} &=\varepsilon^3 (u+w) G_{0 1 0 1 1 2 0 0 0}\, , \\
J_{14} &=\varepsilon^2 (2\varepsilon-1) (3\varepsilon-1) G_{0 1 1 0 0 1 1 0 0}\, , \\
J_{15} &=\varepsilon^3 v G_{0 1 1 0 0 1 2 0 0}\, , \\
J_{16} &=\varepsilon^2 (2\varepsilon-1) (3\varepsilon-1) G_{0 1 1 0 1 1 0 0 0}\, , \\
J_{17} &=\varepsilon^3 u G_{0 1 1 0 1 2 0 0 0}\, , \\
J_{18} &= \frac{1}{25} \varepsilon^2 (1-2\varepsilon)^2 G_{1 0 0 1 1 0 1 0 0}\, , \\
J_{19} &=\varepsilon^2 (2\varepsilon-1) (3\varepsilon-1) G_{1 0 1 0 0 1 1 0 0}\, , \\
J_{20} &=\varepsilon^3 (v+w) G_{1 0 1 0 0 1 2 0 0}\, , \\
J_{21} &= \frac{7}{25} \varepsilon^2 (1-2\varepsilon)^2 G_{1 0 1 0 1 0 1 0 0}\, , \\
J_{22} &= \varepsilon^2 (1-2\varepsilon) m G_{2 1 0 0 0 1 1 0 0}\, , \\
J_{23} &=\varepsilon^3 (u+w) G_{1 1 0 0 0 1 2 0 0}\, , \\
J_{24} &= \frac{7}{25} \varepsilon^2 (1-2\varepsilon)^2 G_{1 1 0 0 1 0 1 0 0}\, , \\
J_{25} &=\varepsilon^4 (v+w) G_{0 0 1 1 1 1 1 0 0}\, , \\
J_{26} &=\varepsilon^4 (u+w) G_{0 1 0 1 1 1 1 0 0}\, , \\
J_{27} &=\varepsilon^4 (u+v) G_{0 1 1 0 1 1 1 0 0}\, , \\
J_{28} &= \frac{1}{2}\varepsilon^2 
\big[-v G_{0 1 0 0 0 2 2 0 0}+m G_{0 1 0 0 2 2 0 0 0}
\\
&\qquad\qquad\qquad\qquad\qquad\quad
+
2\varepsilon v ((u+v+w) G_{0 1 1 0 1 1 2 0 0}-2 G_{0 1 1 0 0 1 2 0 0})\big] \, , \nonumber\\
J_{29} &= \frac{1}{2}\varepsilon^2 
\big[
-2 v G_{0 1 0 0 0 2 2 0 0}-11 m G_{0 1 0 0 2 2 0 0 0}+2\varepsilon u v G_{0 1 1 0 1 2 1 0 0}
\big]\, , \\
J_{30} &= \frac{\varepsilon^2}{v+w}
\big[
-
u v G_{0 1 0 0 0 2 2 0 0} - m (u - 2 v - 2 w) G_{0 1 0 0 2 2 0 0 0}
\\
&
+
u 
\left[
2 \left(6\varepsilon^2-5\varepsilon+1\right) G_{0 1 1 0 0 1 1 0 0}
-
4 \varepsilon v G_{0 1 1 0 0 1 2 0 0} + m v (u+v+w) G_{0 1 1 0 1 2 2 0 0}
\right] 
\big] \, , \nonumber\\
J_{31} &= \frac{1}{2}\varepsilon^2 
\big[
v G_{0 1 0 0 0 2 2 0 0}+12 m G_{0 1 0 0 2 2 0 0 0}+2\varepsilon u (u+v+w) G_{0 1 1 0 2 1 1 0 0}
\big]
\, , \\
J_{32} &= \frac{1}{4}\varepsilon^2 
\big[
v G_{0 1 0 0 0 2 2 0 0}+5 m G_{0 1 0 0 2 2 0 0 0}
\\
&
+
4 
\big[
\left(-6\varepsilon^2+5\varepsilon-1\right) G_{0 1 1 0 0 1 1 0 0} + \varepsilon v G_{0 1 1 0 0 1 2 0 0}
+
\varepsilon m (v+w) G_{0 1 2 0 1 1 1 0 0}
\big]
\big] \, , \nonumber\\
J_{33} &= -\frac{1}{4}\varepsilon^2 
\big[
3 v G_{0 1 0 0 0 2 2 0 0}+17 m G_{0 1 0 0 2 2 0 0 0}
\\
&\qquad\qquad\qquad\qquad\qquad\quad
+
4 \left(6\varepsilon^2-5\varepsilon+1\right) G_{0 1 1 0 0 1 1 0 0}-4\varepsilon m (u+w)  G_{0 2 1 0 1 1 1 0 0}
\big]  \, , \nonumber\\
\label{Element34Eq}
J_{34} &= \frac{1}{5}\varepsilon^2 
\big[ 
\left(14\varepsilon^2-9\varepsilon+1\right) (2 G_{0 1 0 1 1 0 1 0 0} - 4 G_{1 1 0 0 1 0 1 0 0}) +5\varepsilon v G_{0 1 1 1 1 0 1 0 0}
\big]
\, , \\
J_{35} &= (1-2\varepsilon)\varepsilon^3 v G_{0 1 1 1 1 1 0 0 0}\, , \\
J_{36} &=\varepsilon^3 u v G_{0 1 1 1 1 2 0 0 0}\, , \\
J_{37} &=\varepsilon^4 (v+w) G_{1 0 1 0 1 1 1 0 0}\, , \\
J_{38} &= -\frac{7}{5}\varepsilon^3 (2\varepsilon-1) (v+w) G_{1 0 1 1 1 0 1 0 0}\, , \\
J_{39} &=\varepsilon^4 (u+w) G_{1 1 0 0 1 1 1 0 0}\, , \\
J_{40} &= -\frac{7}{5}\varepsilon^3 (2\varepsilon-1) (u+w) G_{1 1 0 1 1 0 1 0 0}\, , \\
J_{41} &= \varepsilon^3 (1-2\varepsilon) u G_{1 1 1 0 0 1 1 0 0}\, , \\
J_{42} &=\varepsilon^3 u v G_{1 1 1 0 0 1 2 0 0}\, , \\
J_{43} &= \frac{1}{5}\varepsilon^2 
\big[
2 \left(14\varepsilon^2-9\varepsilon+1\right) (2 G_{1 0 1 0 1 0 1 0 0}-4 G_{1 1 0 0 1 0 1 0 0} ) + 5\varepsilon u G_{1 1 1 0 1 0 1 0 0} \big]
\, , \\
J_{44} &= \frac{1}{7}\varepsilon^2 
\big[
7 \varepsilon^2v (v+w) G_{0 1 1 1 1 1 1 0 0}
+
2 \varepsilon (7\varepsilon-1) v G_{0 1 1 1 1 0 1 0 0}
\\
&\qquad\qquad\qquad\qquad\qquad\quad
+
2 (2\varepsilon-1) (7\varepsilon-1) (2 G_{0 1 0 1 1 0 1 0 0} - 4G_{1 1 0 0 1 0 1 0 0})
\big] \, , \nonumber\\
J_{45} &= \frac{1}{7}\varepsilon^2 
\big[
\left(14\varepsilon^2-9\varepsilon+1\right) (2 G_{1 0 1 0 1 0 1 0 0}-4G_{1 1 0 0 1 0 1 0 0} ) 
\\
&\qquad\qquad\qquad\qquad\qquad\quad
+ \varepsilon u \big[2 (7\varepsilon-1) G_{1 1 1 0 1 0 1 0 0}+7\varepsilon (u+w) G_{1 1 1 0 1 1 1 0 0} \big]
\big] \, , \nonumber\\
J_{46} &= \frac{1}{5}\varepsilon^2 
\big[
(2\varepsilon-1) (7\varepsilon-1) 
\big[
2 G_{0 1 0 1 1 0 1 0 0}
+
2 G_{1 0 1 0 1 0 1 0 0}
-
8 G_{1 1 0 0 1 0 1 0 0}
\big]
\\
&
+2 \varepsilon (7\varepsilon-1) 
\big[ v G_{0 1 1 1 1 0 1 0 0} + u G_{1 1 1 0 1 0 1 0 0} \big]
-
7 \varepsilon (2\varepsilon-1) u v G_{1 1 1 1 1 0 1 0 0}
\big]
\, . \nonumber
\end{align}
First, we solved the `virtuality' differential equation, in the small-virtuality limit
\begin{align}
\bit{J} = m^{\varepsilon \bit{\scriptscriptstyle A}^{0}_m} \cdot \bit{J}_0 
\end{align}
related to the `massless' MIs $\bit{J}_0$ via the matrix exponent $m^{\varepsilon \bit{\scriptscriptstyle A}^{0}_m}$.
Next, we solved the $m = 0$ limit of the differential equations in Mandelstam-like variables via the Chen iterated integrals on a piece-wise
contour \cite{Chen1977IteratedPI}
\begin{align}
\label{PathIntegralSolution}
\bit{J}_0
=
P_\gamma \exp \left( \varepsilon \int_{[0,u] \cup [0,v] \cup [0,w]} \bit{A}^0 \right) \bit{J}_{00}
\, ,
\end{align}
with the differential of the $A$-matrices $\bit{A} = du \bit{A}^0_u + dv \bit{A}^0_v + dw \bit{A}^0_w$. At each order of the $\varepsilon$-expansion, 
we found solutions in terms of multiple polylogarithms \cite{Goncharov:2001iea}.

%%%%%%%%%%%%%%%%%%%%%%%%%%%%%%%%%%%%%%%%%%%%%%%%%%%%%%%%%%%%%%%%%%%%%
%            Figure
%%%%%%%%%%%%%%%%%%%%%%%%%%%%%%%%%%%%%%%%%%%%%%%%%%%%%%%%%%%%%%%%%%%%%
\begin{figure}[t]
\begin{center}
\mbox{
\begin{picture}(0,60)(237,0)
\put(0,0){\insertfig{16.7}{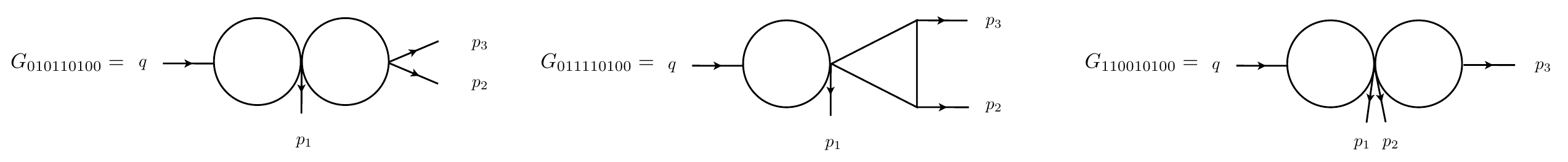}}
\end{picture}
}
\end{center}
\caption{\label{figElement34} Diagrammatic form of the integrals forming the element $J_{34}$ of the canonical basis \re{TriPentCB}.}
\end{figure}
%%%%%%%%%%%%%%%%%%%%%%%%%%%%%%%%%%%%%%%%%%%%%%%%%%%%%%%%%%%%%%%%%%%%%

Finally, we had to fix the vector of the integration constants $\bit{J}_{00}$ at each order of the $\varepsilon$-expansion 
\begin{align}
\bit{J}_{00} = \sum_{p \geq 0} \varepsilon^p \bit{c}^{(p)} 
\, .
\end{align}
To accomplish this, we used two criteria: (i) the absence of spurious poles in the right-hand sides of differential equations at the location of 
$u+v$, $v+w$ and $w+u$ poles and (ii) numerical integration with {\tt FIESTA} \cite{Smirnov:2021rhf} with subsequent use of the {\tt PSLQ} 
algorithm \cite{PSLQ:1999}. However, these considerations alone did no allow us to fully analytically determine all of the integration constants. 
We needed further input. We found that all undetermined contributions are reduced a set of unknowns which can be determined in turn by 
evaluating one of the elements of the canonical basis explicitly. The element in question is $J_{34}$, which is given by a linear combination of 
factorized products of bubbles and triangles, Eq.\ \re{Element34Eq}, as demonstrated in Fig.\ \ref{figElement34}. This can be easily calculated 
making use of the code {\tt MBcreate.m} \cite{Belitsky:2022gba}. It yielded the following expressions
\begin{align}
G_{010110100}
&
=
v^{-\varepsilon} (u+v+w)^{-\varepsilon}
\frac{\e^{2 \gamma  \varepsilon} \Gamma (1-\varepsilon)^4 \Gamma (\varepsilon)^2 }{\Gamma (2-2 \varepsilon)^2}
\, , \\
G_{011110100}
&
=
m^{-2 \varepsilon} v^{-\varepsilon-1}  (u+v+w)^{-\varepsilon} 
\frac{\e^{2 \gamma  \varepsilon} \Gamma (1-\varepsilon)^2 \Gamma (\varepsilon) }{\Gamma (1-2 \varepsilon) \Gamma (2-2 \varepsilon)}
\\
&\times
\big[
m^{2 \varepsilon} \Gamma (-\varepsilon)^2 \Gamma (\varepsilon+1)
+
v^{\varepsilon} \Gamma (1-\varepsilon) \Gamma (\varepsilon) 
\left(
2 m^{\varepsilon} \Gamma (-\varepsilon)+v^{\varepsilon} \Gamma (1-2 \varepsilon) \Gamma (\varepsilon)
\right)
\big]
\, , \nonumber\\
G_{110010100}
&
=
m^{-\varepsilon}  (u+v+w)^{-\varepsilon}
\frac{\e^{2 \gamma  \varepsilon} \Gamma (1-\varepsilon)^4 \Gamma (\varepsilon)^2}{\Gamma (2-2 \varepsilon)^2}
\, .
\end{align}
Matching their expansions to the iterative solution, we found our final result. The expressions are too lengthy to be displayed here in the 
body or appendices, so they are relegated to the accompanying Mathematica notebook {\tt TriPentagonA2Z.nb}, where an interested 
reader could find as well all steps from-A-to-Z for the determination of their expressions starting with necessary initial IBP reductions. The 
tri-pentagon \re{TriPent} is then given by the integral
\begin{align}
{\rm TriPent} = u v (u+v+w) G_{111111100}
\, ,
\end{align}
which is not one of the elements of the above basis, but can be easily reduced to them by means of an IBP reduction. The latter gives
\begin{align}
{\rm TriPent} 
&
=
\frac{1}{\varepsilon^4}
\Bigg[
- J_{3} - \frac{11}{2}  J_{4} 
- 
\frac{2 (1 - 7 \varepsilon) (u + v + w)}{7 (v + w)} [5 J_{11}  - J_{34}]
\nonumber\\
&
\qquad\qquad\qquad\qquad\ \ 
- 
\frac{2 (1 - 7 \varepsilon) (u + v + w)}{7 (u + w)} [5 J_{21}  - J_{43}]
\nonumber\\
&
+
\frac{20 (1 - 7 \varepsilon) (u^2 + 2 u v + v^2 + 3 u w + 3 v w + 2 w^2)}{7 (u + w) (v + w)} J_{24}
\nonumber\\
&
- 
J_{29} + J_{36} + J_{42} + \frac{u}{v + w} J_{44} + \frac{v}{u + w} J_{45} + \frac{5}{7} J_{46}
\Bigg]
\, .
\end{align}
Notice that some of the MIs in this expression do not possess UT individually, however, in the sum ${\rm TriPent}$ is indeed UT. However, 
it is not a pure function: multiple polylogarithms are accompanied by rational prefactors of the $(u,v,w)$ variables.

Before we move on to the next graph, let us point out that there is an alternative way to obtain the analytic result for the tri-pentagon 
\re{TriPent}, which bypasses the need for construction of the MI basis. It is based on the famous triangle relation established in Ref.\ 
\cite{Chetyrkin:1981qh}. Namely, by applying it to the left triangle subgraph in Fig.\ \ref{fig2loop} $(a)$, $G_{111111100}$ 
can be recast in terms of integrals corresponding to graphs with either the central propagator or one of the two (adjacent) left ones 
shrunk into a point. The first of these contributions yields a factorized diagram of a bubble with an attached triangle that can be calculated 
with, say, {\tt MBcreate.m} package \cite{Belitsky:2022gba}. While the second one reduces to the double box integrals evaluated in 
\cite{Belitsky:2023gba} for the required kinematics.

\subsection{Tri-box}

The tri-box graph in Fig.\ \ref{fig2loop} $(b)$ is related to the Davydychev-Ussyukina function $\Phi_2$ given in Eq.\ \re{BoxFunct},
\begin{align}
{\rm TriBox} (p_1, p_2 + p_3)
= (u + w) \Phi_2 (m, v)
\, .
\end{align}
Its small-$m$ expansion immediately produces the sought after expression
\begin{align}
&
{\rm TriBox} (p_1, p_2 + p_3)
=
\left[ \frac{1}{2} \text{Li}_2 \left( \frac{u}{u-1} \right) + \frac{1}{4} \log^2 \frac{1-u}{u} + \frac{\zeta_2}{2}
\right] \log^2 m
\\
&\ \,
+
\left[
3 \text{Li}_3\left(\frac{u}{u-1}\right) - \text{Li}_2 \left(\frac{u}{u-1}\right) \log u - \frac{1}{2} \log (1-u) \log ^2 \frac{1-u}{u}
-
\zeta_2 \log \frac{(1-u)^3}{u^2} 
\right]
\log m
\nonumber\\
&\ \,
+
6 \text{Li}_4\left(\frac{u}{u-1}\right)
-
3 \text{Li}_3 \left(\frac{u}{u-1}\right) \log (u)
+
\frac{1}{2} \text{Li}_2 \left(\frac{u}{u-1}\right) \log ^2 u
\nonumber\\
&\qquad\qquad\qquad\quad\ \
+
\frac{1}{4} \log^2(1-u) \log^2 \frac{1-u}{u}
+
\frac{1}{2} \zeta_2 \log^2 u + 3 \zeta_2 \log (1-u) \log \frac{1-u}{u} + \frac{21}{2} \zeta_4
\, . \nonumber
\end{align}

\subsection{Double boxes and nonplanar tri-box}

The non- and planar double boxes in the near--off-shell kinematics were calculated recently in Ref.\ \cite{Belitsky:2023gba}. The asymptotically
canonical bases for the families of graphs in Fig.\ \ref{fig2loop} $(c)$ and $(d)$ consist of 62 and 97 elements, respectively. Thus, all we need
to do in order to evaluate the integrals in Eqs.\ \re{Dbox}  and  \re{Nbox} is to perform their IBP reduction to the canonical elements constructed 
in \cite{Belitsky:2023gba}. This task is elementary making use of the {\tt FIRE} code and we found
\begin{align}
{\rm DBox}
=
- \frac{1}{2 \varepsilon^4}
\left[
\frac{w}{v} J_{54} + \frac{v}{u+w} J_{55} - J_{59} - J_{61}
\right]
\, ,
\end{align}
and 
\begin{align}
{\rm NBox}
&
=
- \frac{1}{4 \varepsilon^4}
\bigg[
\frac{1585645}{22176}  J_{1} - \frac{949}{528}  J_{2} + \frac{72337}{1848}  J_{3} 
- 
\frac{4403017}{22176}  J_{4} + \frac{7045}{528}  J_{5} - \frac{1153}{88}  J_{8} 
\nonumber\\
&
- 
\frac{1715}{88}  J_{9} - \frac{589 }{44} J_{11} + \frac{3406615}{5544}  J_{12} 
- 
\frac{58561}{11088}  J_{13} - \frac{84}{11}  J_{14} - \frac{1/22} J_{15}  + \frac{6794}{63}  J_{17} 
\nonumber\\
&
+ 
\frac{8495}{126}  J_{18} - \frac{688}{21}  J_{19} - \frac{12673}{3696}  J_{20} 
- 
\frac{42541}{308} J_{21} - \frac{64103}{3696}  J_{22} - \frac{2}{3} J_{23} 
- 
\frac{793}{44}  J_{24} 
\nonumber\\
&
- 
\frac{78637}{308}  J_{25} + \frac{55897}{1848}  J_{26} + \frac{5064085}{22176}  J_{27} 
+  
\frac{799 }{352} J_{28} - \frac{42905}{7392}  J_{29} + J_{33} - 8 J_{34} 
\nonumber\\
&
- 
\frac{5}{2}  J_{35} + \frac{7}{2} J_{36} + 6 J_{37} - 4 J_{38} + 3 J_{39} + 15 J_{40} 
- 
2 J_{41} + \frac{1}{6}  J_{43} - \frac{17 }{6} J_{44} - \frac{71}{21}  J_{45} 
\nonumber\\
&
- 
\frac{82}{7}  J_{46} 
- 
J_{47} + \frac{788}{21}  J_{48} + \frac{464}{21}  J_{49} + 8 J_{50} - 2 J_{54} 
+ 
\frac{5}{2}  J_{55} - \frac{13}{2}  J_{56} - \frac{176}{21}  J_{57} 
\nonumber\\
&
- 
\frac{281}{21}  J_{58} 
- 
\frac{667}{21} J_{59} + \frac{352}{21}  J_{60} - 2 J_{61} - 2 J_{62} + 6 J_{63} 
+ 
11 J_{64} + J_{65} + 7 J_{66} 
\nonumber\\
&
- 
J_{67} + 6 J_{68} + 2 J_{69} + 4 J_{70} 
- 
\frac{122711}{924}  J_{71} - \frac{60253}{462}  J_{72} + \frac{271}{132}  J_{73} 
- 
\frac{1}{3} J_{74} 
\nonumber\\
&
- 
\frac{82953}{308}  J_{75} + \frac{118609}{462}  J_{76} - \frac{17}{44}  J_{77} 
- 
\frac{2 w}{v}  J_{78} + \frac{2 v}{u + w}  J_{79} + \frac{2 w }{u + v} J_{80} 
\nonumber\\
&
- 
\frac{2 (v + w)}{w}  J_{81} 
- 
\frac{3}{11}  J_{82} + \frac{3}{22}  J_{83} - \frac{25}{22}  J_{84} 
+ 
\frac{25}{22}  J_{85} - 4 J_{86} + \frac{25}{22}  J_{87} 
\nonumber\\
&
+ 
\frac{8 (u + v + w)}{v + w} J_{88} + \frac{21}{22}  J_{89} + \frac{1}{44} J_{90} - \frac{25}{132}  J_{91} + \frac{4}{3}  J_{92}
- 
\frac{25}{66}  J_{93}+ \frac{25}{132}  J_{94} - J_{95} + J_{96}
\bigg]
\, ,
\end{align}
for the planar and non-planar graphs in Figs.\ \ref{fig2loop} $(c)$ and $(d)$, respectively. These 
are way too lengthy to be presented in the explicit form in the body of the paper. Therefore, for the reader's convenience, we spell them 
out in the {\tt Mathematica} notebook {\tt Integrals.nb} attached with this submission.

Finally, the nonplanar tri-box in Fig.\ \re{fig2loop} $(e)$ is just one of the MIs in the nonplanar doublebox basis, namely,
\begin{align}
{\rm NTriBox}
=
\frac{2 (u + v + w)}{\varepsilon^4 (v+w)} J_{88}
\, .
\end{align}
Of course, this graph was calculated in Ref.\ \cite{Usyukina:1994iw}, where it was found into factorize after a Fourier transform 
to the square of $\Phi_1$:
\begin{align}
{\rm NTriBox}
=
\frac{1 - u}{2} [\Phi_1(m,u)]^2
\, .
\end{align}
We indeed confirmed our agreement with it on the constraint \re{LDcondition}, $u + v + w = 1 + O(m)$. This concludes our calculation 
of contributing two-loop graphs. All of the integrals reported in this section are UT, however, none are pure.

\section{Adding things up}

Finally, we are in a position to add up all of the calculated integrals.

\label{SumSection}

\subsection{Infrared exponentiation and general structure}
\label{SumSection1}

As we alluded to in the introduction, we anticipate \cite{Bork:2022vat,Belitsky:2022itf,Belitsky:2023ssv} that the infrared logarithms, i.e., 
$\log m$, exponentiate such that the form factor takes the form
\begin{align}
\label{IRExpGen}
\log F_3 = - \frac{\Gamma_{\rm oct}(g)}{4} 
\left[\log^2\left(\frac{m}{u}\right) + \log^2\left(\frac{m}{v}\right)+ \log^2\left(\frac{m}{w} \right)\right]
+
{\rm Fin}_3 \left(u,v,w; g \right)+O(m^2)\, ,
\end{align}
with $\Gamma_{\rm oct}$ being the octagon anomalous dimension \cite{Coronado:2018cxj,Belitsky:2019fan,Basso:2020xts} and ${\rm Fin}_3$ 
being a finite part: it depends only on scalar products of momenta of external states and the 't Hooft coupling constant $g$. It also depends on the 
type of the operator insertion in (\ref{MatrixElement}) as well as helicities of external states. ${\rm Fin}_3$ develops a perturbative expansion
\begin{align}
{\rm Fin}_3 = g^2 f_3^{(1)} + g^4 f_3^{(2)} + \ldots
\, .
\end{align}
The infrared exponent $\Gamma_{\rm oct}$ is known exactly to all orders in the coupling $g$ and is given by \cite{Belitsky:2019fan}:
\begin{align}
\label{ExactGammaOct}
\Gamma_{\rm oct}(g) = -\frac{2}{\pi^2}\log \cosh \left(2 \pi g \right) =  4 g^2 - 16 \zeta_2 g^4  + \ldots\, .
\end{align}
Here, we expanded it to the first two orders, relevant for our current study. We would like to point out the absence in Eq.\ \re{IRExpGen} of linear 
powers in $\log m$ in contrast to the kinematical regime considered in Refs.\ \cite{Alday:2009zm,Henn:2010bk,Henn:2010ir,Henn:2011by}, where 
all external particles' momenta were strictly massless and states propagating in loops' perimeters where taken massive\footnote{The relation 
(\ref{IRExpGen}) is, strictly speaking, a conjecture supported by an array of explicit computations \cite{Bork:2022vat,Belitsky:2022itf,Belitsky:2023ssv} 
as well as a general intuition about IR properties of gauge theories \cite{Mueller:1979ih,Magnea:1990zb,Sterman:2002qn}.}: there is no analogue of
the collinear anomalous dimension in the off-shell regime!

The expansion of $\log F_3$ in powers of $g$ is given by
\begin{align}
\label{logF}
\log F_3=g^2 F_3^{(1)} + g^4 \left( F_3^{(2)}-\frac{1}{2} [F_3^{(1)}]^2\right)+\ldots \, ,
\end{align}
and can be matched onto the expressions for $F_3^{(1)}$ and $F_3^{(2)}$ in terms of scalar integrals given by \re{1loopFF} and \re{2loopFF}, 
respectively. Focusing on the infrared divergent part first, we combine the integrals computed above to find
\begin{align}
\label{logFDiv2loop}
\log F_3\Big{|}_{\rm div}&=
[-3g^2+12\zeta_2g^4+\ldots] \log^2 m + [ 2 g^2 - 8 \zeta_2g^2 + \ldots]\log m \log(uvw)
\, ,
\end{align}
in full agreement with our expectation (\ref{IRExpGen}). 

Several comments are in order. Individual two-loop integrals in \re{2loopFF} contain $\log^4 m$ as well $\log^3 m$ terms. They cancel, however,
in the difference between $F_3^{(2)}$ and the square of one-loop form factor $F_3^{(1)}$ in the $O(g^4)$ coefficient in (\ref{logF}). 
Individual two-loop integrals, i.e., coefficients accompanying the powers of $\log m$, are, in general, expressed in terms of multiple polylogarithms 
\cite{Goncharov:2001iea}. As can be seen in attached {\tt Mathematica} notebook {\tt Integrals.nb}, the coefficients of $\log^2 m$ and $\log m$ in 
(\ref{logFDiv2loop}) are determined solely by ordinary logarithms. To observe the cancellations of higher powers of the infrared logarithms as well 
as simplifications of $\log^2 m$ and $\log m$ terms in (\ref{logFDiv2loop}) we used a combination of the symbol map 
\cite{Goncharov:2009lql,Goncharov:2010jf} along with high-precision numerical computations offered by the {\tt GiNaC} 
integrator \cite{Bauer:2000cp} through the interactive {\tt Ginsh} environment of the {\tt PolyLogTools} package \cite{Duhr:2019tlz}. As we 
emphasized in earlier sections, individual two-loop integrals are not pure UT functions. They, however, do neatly combine into a pure UT 
expression when collected together in $F_3^{(2)}$.

\subsection{Finite part}
\label{SumSection2}

Let us now move on to the finite part ${\rm Fin}_3$. From Eq.\ (\ref{1loopFF}) it is easy to see that at one loop we have
\begin{align}
f_3^{(1)}(u,v,w)=&- \log u \log v  - \log v \log w  - \log w \log u 
\nonumber\\
&
- 2  {\rm Li}_2 (1-u) - 2  {\rm Li}_2 (1-v) - 2  {\rm Li}_2 (1-w)
- 3 \zeta_2 
\, .
\end{align}
The two-loop finite part $f_3^{(2)}$ is given by the $\log m$-free term of the $O(g^4)$ coefficient\footnote{Less $4\zeta_2\left(\log^2(u)+\log^2(v)+
\log^2(w)\right)$ due to our definition of the divergent part which includes a finite term as well.} in (\ref{logF}).  It is a complicated combination 
of multiple polylogarithms of weight~$4$. On the route to simplify this expression, it is instructive to consider its symbol map first 
\cite{Goncharov:2010jf}. Using the {\tt PolyLogTools}, we found out that the symbol of $f_3^{(3)}$ is given by
\begin{align}
\mathcal{S}[ f_3^{(3)} ] = &
-2 u\otimes (1-u)\otimes (1-u)\otimes \frac{1-u}{u}
+u\otimes (1-u)\otimes u\otimes \frac{1-u}{u}
\nonumber \\
&
   -u\otimes (1-u)\otimes v\otimes \frac{1-v}{v}
   -u\otimes (1-u)\otimes w\otimes \frac{1-w}{w}
      \nonumber \\
   &
   -u \otimes v \otimes (1-u)\otimes \frac{1-v}{v}
   -u \otimes v \otimes (1-v)\otimes \frac{1-u}{u}
   \nonumber \\
   &
   +u\otimes v \otimes w \otimes \frac{1-u}{u}
   +u\otimes v \otimes w \otimes \frac{1-v}{v}
   \nonumber \\
   &
      +u\otimes v\otimes w\otimes \frac{1-w}{w}
   -u\otimes w\otimes (1-u)\otimes\frac{1-w}{w}
\nonumber \\
&
   +u\otimes w\otimes v\otimes \frac{1-u}{u}
   +u\otimes w\otimes v\otimes \frac{1-v}{v}
\nonumber \\
&
   +u\otimes w\otimes v\otimes \frac{1-w}{w}
   -u\otimes w\otimes (1-w)\otimes
   \frac{1-u}{u}
   \nonumber \\
& + \ \mathrm{cyclic} \ \mathrm{permutations} \, .
\end{align}
This symbol is identical to the symbol of a local function of the following combination of logarithms and classical polylogarithms:
\begin{align}
\label{ConfRemainder}
R^{(2)}_3(u,v,w)  
= &  -2 \left[ \mathrm{J} \left( -\frac{u v}{w}\right)+\mathrm{J} \left( -\frac{v w}{u}\right)+\mathrm{J} \left( -\frac{w u}{v}\right)\right] 
-
8 \sum_{i=1}^3 \left( \mathrm{Li}_4 \left(1-u_i^{-1}\right)+\frac{\log^4 u_i}{4!} \right) \nonumber\\
& 
-2 \left( \sum_{i=1}^3 \mathrm{Li}_2 (1-u_i^{-1}) \right)^2
+\frac{1}{2} \left( \sum_{i=1}^3 \log^2 u_i\right)^2  -\frac{\log^4(u v w)}{4!}\  , 
\end{align}
with the $\mathrm{J}(z)$ function defined as
\begin{align}
\mathrm{J}(z) = \mathrm{Li}_4(z)-\log(-z) \mathrm{Li}_3(z)+\frac{\log^2(-z)}{2!} \mathrm{Li}_2(z)
-
\frac{\log^3(-z)}{3!} \mathrm{Li}_1(z) - \frac{\log^4(-z)}{48} \ .
\end{align}
Here for brevity of the presentation, we employed the set of variables $u_1\equiv u$, $u_2\equiv v$ and $u_3\equiv w$. The $R^{(2)}_3 $ function 
was first uncovered in the computation of the finite part of the three-gluon form factor in the conformal regime \cite{Brandhuber:2012vm}, i.e., at the 
origin of the moduli space of $\mathcal{N} = 4$ sYM. However, numerical evaluations of $f_3^{(2)}$ and $R^{(2)}_3$ in several kinematical points 
clearly indicate that they are different and the difference is not a constant. This is not surprising given that the symbol map is blind to terms such as 
$\pi^2 \times {\rm function} (u,v,w)$. We have constructed an ansatz of all possible terms\footnote{Taking into account cyclic symmetry as 
well as functional relations between $\mathrm{Li}_2$ reduces the number of terms in the ansatz quite significantly.} of the form $\pi^2 \times \{ \log(x_i)\log(x_j),\, \mathrm{Li}_2(x_i), \, \pi^2  \}$ with rational coefficients plus $R^{(2)}_3$. The values of $x_i$ were taken from the following list
\begin{align}
\left\{
u,v,w,1-u,1-v,1-w,1-\frac{1}{u},1-\frac{1}{v},1-\frac{1}{w}, -\frac{u v}{w}, -\frac{v w}{u},-\frac{w u}{v}
\right\} \ .
\end{align}
Evaluating numerically our ansatz and $f_3^{(2)}$ in several kinematical points using the {\tt Ginsh} integrator allowed us to unambiguously fix 
these coefficients, and we arrived at
\begin{align}
f_3^{(2)}(u,v,w)&
=
R^{(2)}_3(u,v,w)
+
3\zeta_2
\left[\log(u)\log(v)+\log(v)\log(w)+\log(w)\log(u)\right]
\nonumber\\
&
-4\zeta_2\sum_{i=1}^3\mathrm{Li}_2\left(1-u_i^{-1}\right)
+
9
\zeta_2\sum_{i=1}^3\log^2 u_i
+
\frac{63\zeta_4}{4}\,  . 
\end{align}
This concludes our calculation of the finite part at the two-loop order. We see that it is a pure function of uniform transcendentality just 
as in the conformal case.

\subsection{Iterative structure}
\label{SumSection3}

In the massless case of scattering amplitudes, it became customary to split results according to the so called BDS ansatz 
\cite{Bern:2005iz} and a finite remainder \cite{Bern:2008ap,Drummond:2008aq}. The same decomposition was established for the case of 
form factors as well \cite{Brandhuber:2012vm}. Such a decomposition admits the following generic from
\begin{align}
F_3^{(2)} = \frac12 \big[ F_3^{(1)} \big]^2 + 4 \zeta_2 \widetilde{F}_3^{(1)} + \mathcal{R}_3^{(2)}
\, .
\end{align}
In the massless case, $\widetilde{F}_3^{(1)}$ was found to enjoy a very powerful feature, namely, it was determined at two loops to be merely 
given by the one-loop form factor \cite{Brandhuber:2012vm} 
\begin{align}
\widetilde{F}_3^{(1)} = \frac14 F_3^{(1)}  (2 \varepsilon)
\, ,
\end{align}
where the factor of $\ft14$ is introduced to accommodate the change from $\Gamma_{\rm oct}$ to $\Gamma_{\rm cusp}$ of the massless case. 
This is the well-known cross-order relation \cite{Bern:2005iz} encoding the iterative structure of massless amplitudes. It was also confirmed on the 
Coulomb branch where the external legs were kept massless \cite{Henn:2010ir}.

Adopting the same nomenclature in the current `off-shell' case, we find that $\widetilde{F}_3^{(1)}$ possesses all of the building blocks of the 
one-loop form factor $F_3^{(1)} $ but is not directly related to it except for the infrared-divergent terms. It has the form 
\begin{align}
\widetilde{F}_3^{(1)} 
&= 3 \log^2 m - 2 \log m \log u v w \\
&+ \frac34 \big[ \log^2 u + \log^2 v + \log^2 w + \log u \log v + \log u \log w + \log v \log w \big] \nonumber\\
&+ {\rm Li}_2 (1 - u) + {\rm Li}_2 (1 - v) + {\rm Li}_2 (1 - w) \nonumber
\, ,
\end{align}
cf.\ Eq.\ (\ref{logFDiv2loop}), such that $\widetilde{F}_3^{(1)} |_{\rm div} = F_3^{(1)} |_{\rm div}$. With this convention, the `off-shell' remainder 
function $\mathcal{R}_3^{(2)}$ is related by a constant shift
\begin{align}
\mathcal{R}_3^{(2)}(u,v,w)=R^{(2)}_3(u,v,w) + \frac{63\zeta_4}{4}\,  .
\end{align}
to the one of the conformal case\footnote{Note that the remainder function in Ref.\ \cite{Brandhuber:2012vm} contains an additive constant
$-\ft{23}{2} \zeta_4$, which we did not include in our definition \re{ConfRemainder}.}, $R_3^{(2)}$  \cite{Brandhuber:2012vm}! Indeed, we 
could enforce the same iterative structure of the `off-shell' form factor as in the conformal case at the expense of changing the remainder
function $\mathcal{R}_3^{(2)}$.

\section{Conclusion}

With this paper, we continued our excursion into the land of the Coulomb branch away from the origin in its moduli space.
The object under our study was form factor of the lowest component of the stress-tensor multiplet for three massive 
W-bosons. We were particularly interested in the asymptotic region of their vanishing masses, $m \to 0$. In this case, the 
emerging infrared divergences are encoded by the logarithms of $m$, which replace inverse powers of $\varepsilon$
in dimensional regularization. However, this is not to be confused with another use of the Coulomb branch advocated in 
Ref.\ \cite{Alday:2009zm}, as a means to make amplitudes and form factors finite by giving vacuum expectation values to 
scalars propagating around quantum loops perimeters. In the latter case, it was established that amplitudes and Sudakov 
form factors echo the well-known infrared behavior of massless scattering amplitudes and form factors with the infrared
physics driven by the cusp anomalous dimension. In counter-distinction, we find instead, that like in the case of  scattering 
amplitudes of four- \cite{Caron-Huot:2021usw} and five W-bosons \cite{Bork:2022vat} and the Sudakov form factor of two 
W-bosons \cite{Belitsky:2022itf,Belitsky:2023ssv}, the infrared logarithms are accompanied by a completely different function of the 
coupling, the octagon anomalous dimension  \cite{Coronado:2018cxj,Belitsky:2019fan,Basso:2020xts}. This reconfirms
the role of the latter as the critical infrared exponent of the off-shell kinematics.

Further, the form factor of three W-bosons possesses a nontrivial remainder function. After a proper subtraction of infrared
logarithms with judiciously-chosen finite parts, we found it to be identical to the one in the massless case (up to a constant), 
i.e., the origin of the moduli space. The structure of the collinear limit is however quite different in the two cases. While the 
massless case inherits its iterative structure in terms of one-loop amplitude/form factor, the case of massive W-bosons is trickier. 
In order to put it on a firmer foundation, analysis of the five-W amplitude at generic values of Mandelstam-like variables needs to be
studied, as opposed to the symmetric point discussed in Ref.\ \cite{Bork:2022vat}. 

Last but certainly not least is the question of  the dual description of scattering amplitudes and form factors on the Coulomb 
branch. A proposal for an off-shell Wilson loop was put forward in Ref.\ \cite{Belitsky:2021huz} starting from a higher-dimensional 
holonomy and dimensionally reducing it down to four-dimensions. However, while the one-loop expectation value for four sites 
was found to be in agreement with the amplitude of the W-bosons, starting from two loops the two `observables' started to deviate. 
The reason for this fact remains obscure. The T-dual gauge theory was chosen to be the conformal $\mathcal{N}=4$ sYM. Had it
rather be something else or one had to use a different variant of dimensional reduction? This question will have to be readdressed 
in the future.

\begin{acknowledgments}
We would like to thank Johannes Henn, Roman Lee and Alexander Smirnov for useful communications and discussions. 
The work of A.B.\ was supported by the U.S.\ National Science Foundation under the grant No.\ PHY-2207138. The work 
of L.B.\ was supported by the Foundation for the Advancement of Theoretical Physics and Mathematics “BASIS”. The work of 
V.S.\ was supported by the Russian Science Foundation under the agreement No.\ 21-71-30003 (solving differential equations
for Feynman integrals) and by the Ministry of Education and Science of the Russian Federation as part of the program of 
the Moscow Center for Fundamental and Applied Mathematics under Agreement No.\  075-15-2022-284 (applying {\tt FIESTA} 
for checks and to obtain precision values for integration constants).
\end{acknowledgments}

%%%%%%%%%%%%%%%%%%%%%%%%%%%%%%%%%%%%%%%%% 

\end{document}